\documentclass[12pt,preprint]{emulateapj}	
\usepackage{epsfig}
\usepackage{floatflt}
\submitted{Submitted to The Astrophysical Journal on October 20, 2010} 
\begin{document}                           			
\title{The Merger Environment of the WAT Hosting Cluster Abell 562}  							
\author{E.M. Douglass\altaffilmark{1},
            Elizabeth L. Blanton\altaffilmark{1},
            T.E. Clarke\altaffilmark{2},
            Scott W. Randall\altaffilmark{3},
            Joshua D. Wing\altaffilmark{1}}

 	\altaffiltext{1}{Institute for Astrophysical Research, Department of Astronomy, Boston University, Boston, MA 02215}
    	\altaffiltext{2}{Naval Research Laboratory, Washington, DC. 20375}
	\altaffiltext{3}{Harvard-Smithsonian Center for Astrophysics, 60 Garden Street, Cambridge, MA 02138}

\begin{abstract}
We present a \emph{Chandra} X-ray observation and VLA radio observations of the nearby (z=0.11) galaxy cluster Abell 562 and the wide angle tail (WAT) radio source 0647+693.  The cluster displays signatures of an ongoing merger leading to the bending of the WAT source including an elongation of the X-ray surface brightness distribution along the line that bisects the WAT, an excess of displaced gas found between the radio lobes, and anisotropies within the ICM projected temperature and abundance distributions.  The most likely geometry of the ongoing interaction is a head-on merger occurring along the WAT bending axis.  By combining observable properties of A562 and 0647+693 with common values for the conditions within merging clusters at the time of core crossing, we constrain the internal density ($\rho_j$=0.001 $\rho_{ICM}$) of the jets and plasma flow velocity within the lobes (v$_r$ = 0.02c - 0.03c) of the WAT source.        
\end{abstract}


\section{INTRODUCTION}  
Wide angle tail (WAT) radio sources are high power FR-I type sources (P$_{1.4}$$\sim$ 10$^{25}$ W Hz$^{-1}$) generally associated with the bright central galaxy at or near the center of its host cluster.  WATs contain twin, well collimated jets extending anywhere from tens to hundreds of kpcs from the center of the host galaxy.  The jets flare abruptly at bright hotspots beyond which extend edge-darkened lobes, often bent in a similar direction.  It is this bending that led to the name \emph{wide angle tail} when first classified by \citet{OR76}.  The WAT definition has since been expanded to include any cluster center source of intermediate radio power which displays the aforementioned jet$/$hotspot$/$lobe transition, regardless of whether it appears bent \citep{Hardcastle98}.  
\par
The bending of WATs was originally believed to result from intracluster medium (ICM) ram pressure induced by the motion of the host galaxy relative to a relaxed ICM.  Early modeling of WATs showed that significant ICM ram pressure (v$_{gal}$ $>$ 1000 km s$^{-1}$) is necessary to shape many of the bent sources (Eilek et al. 1984, O'Donoghue et al. 1993).  Low peculiar velocities (v$_{g}$ $<$ 300 km s$^{-1}$) of the large central dominant galaxies (cD,D,gE) with which WATs are typically associated \citep{Burns81} argue against peculiar galaxy motion as the WAT bending mechanism.  Therefore, cluster mergers have been invoked as the primary scenario for WAT bending \citep{Burns94}.  Relative velocities between two merging systems can easily exceed 1000 km s$^{-1}$ resulting in significant bulk flow of the ICM with respect to cluster galaxies.  Simulations have shown that, within the central 200 kpc of the cluster, ICM flow velocities may remain above 1000 km s$^{-1}$ for timescales much longer (t$_{\rm flow}$ $\sim$ 10$^9$ yr) than the typical lifetimes of the radio sources themselves (t$_{\rm rad} \sim$ 10$^7$-10$^8$ yr, \citet{Loken95,Roettiger96}).  From an observational standpoint, elongations in the ICM X-ray surface brightness distribution along the line that bisects the WAT \citep{Gomez97}, the significant offset ($\ge$ 100 kpc) of the WAT from the X-ray centroid \citep{SM00}, and the presence of ICM substructure \citep{Burns94, Gomez97}, have all been observed in WAT systems and are believed to be indicators of recent cluster mergers.
\par
Large velocities relative to the ICM resulting from cluster mergers may not be required to bend the WAT if a model of high flow velocity and low density is assumed within the radio lobes \citep{BB82, Eilek84, ODon90, Pinkney95, Hardcastle05}.  \citet{Hardcastle05} use this model to argue that galaxy velocities as low as 100-300 km s$^{-1}$ are sufficient to produce the WAT morphology in some clusters. In previous models as mentioned above, denser lobes with lower flow velocities require galaxy velocities greater than 1000 km s$^{-1}$.  

Their large radial extents, significant radio powers, and association with galaxy clusters allows for the use of WATs as tracers of high redshift systems, where X-ray and optical observations are more difficult \citep{Blantonet00}.  If we are to use WAT clusters at high redshift for studies of cosmology and galaxy evolution, it is important to understand the conditions found within them and to determine whether they differ from clusters identified using other methods.

Here we report on the results of a 51 ks \emph{Chandra} observation of the WAT host cluster Abell 562 at a redshift of z=0.11.  The cluster has an Abell Richness Class of 1.  The bright, central, large elliptical galaxy is host to the significantly bent WAT source 0647+693.  Abell 562 has been previously observed with the \emph{ASCA} (PI: Hanami) and \emph{ROSAT} \citep{Gomez97} observatories.  Unless otherwise noted, all uncertainty ranges are 90\% confidence intervals.  We assume $\Omega _{\Lambda} =0.7$, $\Omega _{M}=0.3$, and $H_{\rm o}=70$ km s$^{-1}$ Mpc$^{-1}$.  At $z=0.110$, the angular diameter distance is $D_{\Theta}=413.70$ Mpc, the luminosity distance is $D_{\rm L}=509.71$ Mpc, and $1\arcsec = 2.0$ kpc. \\

\section{OBSERVATION AND DATA REDUCTION}
\subsection{\emph{Chandra} Observation}
Abell 562 was observed by \emph{Chandra} on 2006 December 1 for a total of 51,611 seconds in very faint (VFAINT) mode.  The center of the cluster was positioned on the S3 chip of \emph{Chandra}'s ACIS-S detector with a 1$\arcmin$  offset from the nominal pointing to avoid node boundaries. The data analysis package CIAO version 4.2 was used for data reduction.  By analyzing light curves of the S1 and S3 chips, it was determined that there were no strong background flares. The data were corrected for hot pixels and cosmic ray afterglows using standard techniques. The cleaned level 2 events file had an exposure time of 51,467 seconds.  Period E background files were taken from the blank sky observations of M. Markevitch included in the CIAO calibration database (CALDB), cleaned  for VFAINT mode, and reprojected to match the A562 observation. Count rates within similar regions in images restricted to the 10-12 keV energy range were compared to determine the normalization of the background files.    

\subsection{Radio Data}
Radio observations of 0647+693 at frequencies near 1.4 GHz (PI: Lacey) and 8.4 GHz (PI: Hardcastle) were extracted from the NRAO VLA archive. The 8.4 GHz data were presented in \citet{HardSak04}.  The 1.4 GHz data (taken in spectral line mode) targeted a nearby VLA calibrator source, 3C169, which is located $\sim$ 11\farcm5 away from 0647+693 while the 8.4 GHz data (taken in continuum mode) targeted the source of interest. The observing parameters for the radio data are provided in Table~\ref{tbl:radio}.

The 1.4 GHz observations were taken in the VLA's most extended A configuration and used 3C48 as a flux calibrator. These data were calibrated and reduced in spectral line mode to reduce the effects of bandwidth smearing at the location of the target. Using the NRAO Astronomical Image Processing System (AIPS), the images were produced through the standard Fourier transform deconvolution method. The data were processed through several loops of imaging and self-calibration to reduce the effects of phase and amplitude errors. The final images were divided by VLA's primary beam response to correct the flux at the position of 0647+693.

Observations at 8.4 GHz were extracted for the VLA's A, B, and D configurations to produce images sensitive to a wide range of structures. All three configurations used 0607+673 as a phase calibrator. The flux calibrator for the A and B configuration data was 3C48, while the D configuration data were flux calibrated with 3C286. Each configuration was calibrated, imaged, and self-calibrated separately using standard techniques in AIPS. We have created a high resolution image ($\theta$$_{FWHM}$=0.87 $\times$ 0.68 arcsec) for study of the jet structure by combining the A and B configuration data and a lower resolution image ($\theta$$_{FWHM}$=1.9 $\times$ 1.7 arcsec) sensitive to the diffuse extended lobes by combining all three configurations.

To study the spectral index structure of the target we have cut the uv coverage of the 8.4 GHz low resolution dataset so that it matches the 1.4 GHz data. We then convolved the (nearly identical resolution) 1.4 and 8.4 GHz images to a circular beam of $\theta$$_{FWHM}$ = 4.26\arcsec. Finally, the two input maps were blanked at the three sigma level on each map prior to constructing the spectral index map of the target.

\subsection{Optical Data}
Optical observations of Abell 562 were carried out on 22 and 23 April 2009 with the Perkins 1.83 m telescope at Lowell Observatory using the PRISM instrument in imaging mode \citep{Janes04}.  The cluster was observed in the R-band for a total of 3600 sec with an average seeing $<$ 2$\arcsec$. The data were reduced using standard \emph{IRAF} procedures for bias and flat-correction.  Twilight flats were used in the reduction.   

\section{SPATIAL DISTRIBUTION OF CLUSTER X-RAY EMISSION}
An adaptively smoothed 0.3-7.0 keV \emph{Chandra} X-ray image of the cluster was created using the \emph{csmooth} task within CIAO with a minimum SNR of 3, and is shown in Fig. 1.  It has been corrected for background and exposure.  The surface brightness distribution is significantly elongated along the line that bisects the WAT (Fig. 3).  In addition to the elongation there appears to be substructure within the X-ray emission.  This includes an excess near the central galaxy, running roughly perpendicular to the elongation, and a decrement coincident with the northern radio lobe.    

Individual sources were detected using the wavelet detection algorithm \emph{wavdetect} in CIAO. Of the 23 sources that were detected, five correspond with sources in the USNO A2.0 catalog.  The brightest central galaxy (WAT host, R=15.5) and its next brightest neighbor (R=15.7), 20$\arcsec$ to the east, were both identified as X-ray sources.  The five \emph{wavdetect} sources were identified within 1$\arcsec$ of their optical counterparts suggesting that there is no positional offset of the X-ray image.  Therefore no astrometric correction has been applied to the data.  These sources can be seen as the red squares in the optical image of the cluster (Fig. 2).  The third brightest galaxy (R=16.1), positioned 150$\arcsec$ to the east of the central bright pair, was not detected in the X-ray.  All the X-ray point sources detected were excluded from the analysis of the extended emission.    

\subsection{ICM and WAT Interaction}
The central elliptical galaxy is host to the WAT radio source 0647+693 (Fig. 3).  The source consists of two bent jets that extend a projected distance of $\sim$ 80 kpc after which they disrupt within their edge-darkened lobes.  Near the jets, the lobes appear to bend along a similar northeast-southwest direction, presumably due to the ram pressure resulting from ICM flow parallel to the surface brightness elongation.  At a radius of 300 kpc, the lobes bend $\sim$ 90$^\circ$ to point radially outward from the cluster center.  This is likely the radius at which buoyancy overtakes ram pressure and can be used to constrain the ICM flow velocity (see $\S$  6.2).  At a radius of 350 kpc from the cluster center, the northern lobe displays an additional bend, appearing anti-parallel to the inner lobe's ICM ram pressure orientation.  This may result from merger induced turbulence within the gas.  

\subsection{Surface Brightness Distribution}
We fit a two-dimensional elliptical $\beta$-model to a 0.3 - 7.0 keV image of the cluster with 16 x 16 pixel binning using the \emph{beta2D} model within \emph{Sherpa4.2}.  We determined a core radius of $R_c = 80.239^{+11.17}_{-10.14}$ kpc, with $\beta$ = 0.34 $\pm$ 0.013, e=0.27$\pm$0.03, and P.A. = 60$^\circ$ $\pm$ 4.2$^\circ$ (measured from north to east).  The center of the large scale emission was determined to be located 7$\arcsec$ to the southeast of the AGN.  These values are close to those obtained in \citet{Gomez97}, of $R_c = 99 \pm 9$ kpc and $\beta = 0.5 \pm 0.3$,  and are typical for poor clusters (Mulchaey 2000).  

In an attempt to further characterize the X-ray surface brightness distribution we fit a double $\beta$-model to the data consisting of two single \emph{beta2D} models with linked ellipticities, position angles, and centers.  The fit parameters were determined to be: $R_{c1}$ = 80.65 kpc, $\beta_{1}$ = 0.34 , $R_{c2}$ = 26.06 kpc, $\beta_{2}$ = 9.9, e=0.27, and P.A. = 60$^\circ$ with an amplitude ratio of $\sim$ 1.6:1.  The errors on the values were poorly constrained with many reaching the hard limit.  Therefore the single \emph{beta2D} model was considered to be the better fit.

\subsection{Cluster Substructure}
It is clear from the adaptively smoothed image of the ICM emission (Fig. 1) that there is noticeable substructure within the central 150$\arcsec$ (300 kpc) of the cluster.  
To better show the substructure we subtracted the 2-D $\beta$-model ($\S$ 3.2) from the X-ray image.  As can be seen in Fig. 4, an excess of emission is located between the radio lobes directly to the southwest of the cluster center.  This excess is most likely gas displaced due to ram pressure by an infalling subcluster (see $\S$ 6.1).

\section{SPECTRAL ANALYSIS}
\subsection{Total Spectrum}
A spectrum was extracted from an elliptical region centered on the AGN with a semi-major axis of 200$\arcsec$ (400 kpc), e = 0.27, and P.A. = 60$^{\circ}$, aligned with the best-fit 2-D $\beta$-model from $\S$ 3.2.  The central AGN and all other X-ray point sources were excluded.  The spectrum was binned such that each energy bin contained a minimum of 40 counts.  Restricting the energy range to 0.7-7.0 keV, an absorbed single temperature APEC model was fitted to the spectrum using XSPEC V12.6.0. Temperature, abundance, and normalization were free parameters.  The absorption parameter was fixed at the Galactic value of 5.48 $\times$ 10$^{20}$ cm$^{-2}$ \citep{DL90}.  Additional models were also fitted including a second APEC model and a cooling flow model (MKCFLOW).  The lack of an improved fit with the addition of these components is consistent with the cluster ICM being comprised of nearly isothermal gas.  Determined from $\sim$ 67,000 source counts, the average temperature of the ICM was determined to be 2.76$^{+0.21}_{-0.14}$ keV with an average abundance value of 0.20 $^{+0.08}_{-0.07}$ Z$_{\odot}$ from the single APEC fit with fixed Galactic absorption. The fit was good with a $\chi^2$ of 268.0 for 313 degrees of freedom, resulting in a reduced $\chi^2$ statistic of 0.86.

\subsection{Temperature and Abundance Profiles}
A temperature profile was created by extracting the spectra from 15 concentric elliptical annuli of the same ellipticity and position angle as above ($\S 4.1$).  Each annulus contained at least 1000 source counts.  The spectra were fitted with an APEC model with temperature and normalization allowed to vary.  Abundance was fixed at the global value.  As can be seen in Fig. 5, the temperature remains roughly constant ($\sim$3.0 keV) out to a radius of 400 kpc (200$\arcsec$). An abundance profile was extracted using two concentric elliptical annuli.  As can be seen in Fig. 5, the profile values and uncertainties restrict the abundance to the range 0.02-0.5 Z$_\odot$ in the two annuli.  While the profile for the chemical abundance hints at a slight decrease with increasing radius, the decrease is not statistically significant.  The values of temperature and chemical abundance determined from the radial profiles agree well with the average values determined from the best-fitting single-temperature APEC model of the total cluster spectrum.

\subsection{Temperature and Abundance Map}
A 40 x 40 pixel projected temperature map was created of a region 500 kpc (250$\arcsec$) on a side centered on the AGN (Fig. 6a).  Each value in the temperature map was determined by extracting a spectrum from the smallest surrounding region on the chip containing 1000 total counts.  Each spectrum was fitted with a single temperature APEC model with the temperature, abundance, and normalization as free parameters.  The uncertainties of the temperature values in the map are roughly 15\% at the center, to 40\% towards the edges.  As seen in the temperature profile, values of temperature remain relatively constant when averaged over annuli.  However, the temperature map reveals temperature variations within the core of the cluster.  Lying just east of the AGN, a region of increased temperature stretches roughly 60$\arcsec$ N-S (120 kpc) and 40$\arcsec$ E-W (80 kpc). To more accurately assess the thermal characteristics, spectra were extracted from an elliptical region of the aforementioned dimensions and a larger surrounding elliptical region.  The enhancement was determined to be significantly hotter than its surroundings, with their temperatures being kT = 3.43$^{+0.54}_{-0.39}$ keV and kT = 2.48$^{+0.29}_{-0.27}$ keV respectively.  It is possible that such a temperature enhancement is indicative of a merger shock. This possibility will be discussed further in $\S 6.1$. 

The companion 40 x 40 pixel abundance map of A562 is presented in Figure 6b.  While the uncertainty in the values is relatively high (40\% towards the center and 80\% towards the edges), the presence of higher abundance gas located between the WAT lobes is intriguing. A spectrum was extracted from a circular region of radius 30$\arcsec$ centered on the abundance enhancement. A second spectrum was extracted from a circular region of radius 85 $\arcsec$, centered on the AGN, with the region of interest and point sources excluded.  Both spectra were fitted with an \emph{APEC} model in XSPEC accounting for Galactic absorption.  The value of the abundance between the lobes was determined to be Z=0.96$^{+1.12}_{-0.44} Z_{\odot}$ while the surrounding region was Z=0.25$^{+0.11}_{-0.10} Z_{\odot}$.  This high abundance gas lies within the excess emission revealed in the $\beta$-model subtracted image of the cluster (Fig. 4) and will be discussed in $\S 6.1$.

\subsection{$L_{X}-T$ Relation}
Following the method described in \citet{Wu99}, we determined the total bolometric luminosity of Abell 562.  The bolometric luminosity of this region was calculated from the total cluster spectrum discussed above accounting for redshift and cosmology (following \citet{Wu99}, H$_0$ = 50 km s$^{-1}$ Mpc$^{-1}$ and $\Omega_m$ = 1).  A `dummy' response file was created to allow the spectral model to be evaluated from 0.001-100 keV.  As the \emph{Chandra} observation of A562 only extends out to $\sim$400 kpc (200$\arcsec$) from the cluster center,  the contribution of the ICM emission not included on the S3 chip had to be estimated.  Given that the $\beta$ value within the surface brightness fits discussed in $\S 3.2$ is rather shallow ($\beta$=0.34), an unrealistic, non-zero surface brightness is returned when extrapolated out to large radii. While our model is appropriate to use for the regions of the cluster covered by the S3 chip, we use the \citet{Gomez97} value $\beta$ = 0.5 for r $>$ 400kpc.  From this analysis it appears that A562, given its global temperature, has a luminosity falling along the $L_{X}-T$ trend when compared with the \citet{Wu99} sample.  This can be seen in Fig. 7.  From the simulations of \citet{RT02}, clusters undergoing major mergers have boosts in their X-ray luminosity and temperature such that they follow the slope of the relation, but generally fall above the trend line throughout the duration of the merger.  As seen in Fig. 7, Abell 562 falls along the trend.  Corrected for our cosmology, the total bolometric luminosity of the cluster is L$_X$ = 8.55 $\times$ 10$^{43}$ ergs s$^{-1}$.

\section{Pressure, Density, Cooling Time, and Mass}
\subsection{Pressure and Density Profiles}
Assuming that the emissivity is constant in elliptical shells, a routine that deprojects the X-ray surface brightness (see \citet{Blanton09}) was used to give the emissivity as a function of radius. Pressure and density profiles were then calculated using the temperatures from $\S$4.2.  As can be seen in Fig. 8, density and pressure generally decrease with increasing radius.  From a radius of roughly 400 kpc (200$\arcsec$), the density increases by a factor of five to a core density of 3.00$\pm 0.40$ $\times$ 10$^{-3}$ cm$^{-3}$.  The gas pressure also increases by a factor of five from a radius of 400 kpc (200$\arcsec$) to its core.  The pressure reaches a value of 3.25$^{+0.64}_{-0.83}$ $\times$ 10$^{-11}$ dyn cm$^{-2}$ near the cluster center. There is an apparent pressure jump between $\sim$ 70$\arcsec$ and $\sim$ 90$\arcsec$, possibly indicating the presence of a merger induced shock.  Using Rankine-Hugoniot jump conditions at this pressure jump (assuming the maximum value within the uncertainty from the post-shock annulus and minimum value for the pre-shock annulus), we determine an upper limit on the Mach number to be \emph{M}=1.6. Given the temperature at this radius kT = 3.77$^{+0.83}_{-0.68}$ keV with a corresponding sound speed of v$_s$ = 687 km s$^{-1}$, we estimate a bulk flow velocity of v $\sim$ 1100 km s$^{-1}$, consistent with what might be expected within an ongoing merger.         

Using \emph{ROSAT} data, \citet{Gomez97} found the pressure near the center to be 1.9 $\times$ 10$^{-12}$ dyn cm$^{-2}$, significantly lower than the value obtained with \emph{Chandra}.  This difference may be a result of the larger core radius $\beta$-model from which they determined density, along with their lower global cluster temperature (kT $\sim$ 1.7$^{+1.7}_{-0.4}$ keV).  A similar discrepancy between \emph{ROSAT} and \emph{Chandra} derived pressures is seen with the WAT clusters A1446 \citep{Douglass08} and A2462 \citep{Jetha05}.

\subsection{Cooling Time}
Using the temperature and density profiles determined above ($\S 5.1$), we calculated the cooling time of the intracluster gas (Fig. 8) following the equation from Sarazin (1988):
\begin{equation}       
t_{cool}=8.5\times10^{10} yr (\frac{n_p}{10^{-3}cm^{-3}})^{-1}(\frac{T_g}{10^8 K})^{1/2},     
\end{equation}

From the temperature profile it appears that temperature does not decrease with decreasing radius.  The cooling time of the central ICM is never below the Hubble time, with a value of $\sim$ 15 Gyr within the central annulus.  Therefore, we conclude that A562 is likely not the site of an ongoing cooling flow.

\subsection{Gas and Gravitational Mass}
The gas and gravitational mass were determined using a routine (as in $\S 5.1$) that deprojects the X-ray surface brightness of the extended emission to give emissivity and density as a function of radius.  These are integrated to give the enclosed gas mass as a function of radius.  The total gravitational mass was derived from the equation of hydrostatic equilibrium.  As can be seen in Fig. 9, the gas mass fraction (ratio of gas mass to gravitational mass, GMF) at the cluster core is  $\sim$0.009 which increases with radius to $\sim$0.02 at 400 kpc (200$\arcsec$) which is significantly lower than the typical value for the GMF for galaxy clusters at this radius \citep{Allen04}.  This suggests that A562 deviates from hydrostatic equilibrium, as is typical of clusters undergoing significant mergers \citep{PB07,Poole06,Cavaliere99}.

\section{DISCUSSION}
\subsection{Cluster-Cluster Merger}
\subsubsection{Merger Geometry} 
Previous X-ray studies of WAT clusters have revealed that many bent sources are associated with clusters exhibiting classic merger characteristics, including: elongation of the ICM distribution along the line that bisects the WAT \citep{Gomez97}, significant offset of the WAT-hosting bright central galaxy from the X-ray centroid \citep{SM00}, and X-ray substructure within the cluster core \citep{Burns94}.  As WATs show a broad range of bending angles, including no bend, it has been argued that the formation of WATs may not be directly linked to cluster dynamical activity \citep{Hardcastle05,Jetha06}.  Instead, it is possible that those WATs which show significantly bent jets are found at the site of ongoing mergers, while those WATs with straight jets are situated in more relaxed environments.  From \emph{ROSAT} observations of its disturbed ICM environment and its strongly bent well-collimated jets, 0647+693 is often considered the prototypical merger-induced WAT.  

As can be seen in Figure 3, the surface brightness distribution of Abell 562 exhibits a significant elongation parallel to the line which bisects the opening angle of 0647+693.  This is consistent with some component of a merger currently occurring along this axis.  Simulations of cluster mergers have shown that ICM flow within the cores of clusters ($\sim$ 200 kpc) may remain above 1000 km s$^{-1}$ for t $>$ 2 Gyr \citep{Roettiger96}.  Such an ICM bulk flow velocity would result in the bending of the jets and inner lobes into their observed shape (see $\S$  6.2).  

During a head-on merger it is expected that the majority of the most observable merger-related features in the X-ray will lie along the collision axis.  Within the temperature map, a region of increased temperature is revealed along this presumed axis to the ENE of the AGN.  The region is coincident with what appears to be a surface brightness edge in the adaptively smoothed image of the cluster.  Such a feature is likely an indication of a merger shock created by an infalling subcluster from the east. As seen in the simulations of \citet{Paul10}, during a merger, at the moment of core crossing, forward and reverse shocks can propagate outward from the cluster center along the merger axis. In an effort to better characterize this apparent density discontinuity, we followed the method outlined in \citet{Randall09}. Profiles of partial concentric annuli, centered on the AGN, were fitted with a spherical gas density model consisting of two power laws, with jump amplitude, radius, normalization, and inner and outer slopes as free parameters. The results were inconclusive, likely due to the non-spherical symmetry of the feature and the faintness of the shock edge.  

As seen in Figure 4, there is an excess of emission lying between the WAT lobes and extending $\sim$ 50$\arcsec$ (100 kpc) southwest of the AGN.  This is most likely previously centrally located gas displaced by an infalling subcluster.  This excess again lies along the presumed merger axis, with its location consistent with a head-on collision.  In the abundance map (Fig. 6b), at the location of the excess emission, there is an apparent region of significantly increased metallicity (see $\S$ 4.3).  Using the normalization of the spectral fit for this region, which we assume to be a sphere, we estimate the mass of the high-metallicity gas to be M $\sim$ 5 $\times$ 10$^{10}$ M$_{\odot}$. As this value is too large to simply be the stripped halo of the host galaxy, it is more likely that this is the remnant cool core of the pre-merger cluster which had formerly been coincident with the WAT host.  The possibility exists that this excess may instead be gas evacuated from the northern cavity, but the high metallicity of the region suggests against this interpretation.   

Many similar cool core remnants were presented in Rossetti \& Molendi (2009).  Within this scenario, the cool gas was possibly reheated by the merger.  Within \citet{Gomez02}, it is argued that the disruption of a cooling flow may occur when the ram pressure of the infalling gas exceeds that of the thermal pressure of the cool core.  Assuming the central pressure of the pre-merger core to be similar to that of the weak cool core in the WAT host cluster A2634 \citep{Hardcastle05}, where P$\sim$10$^{-10}$ dyn cm$^{-2}$, the merger induced ICM bulk flow velocity would need to exceed $\sim$ 1000 km s$^{-1}$ to disrupt the core (see $\S$  6.2).  As mentioned earlier, such velocities are expected within the cores of merging clusters up to hundreds of millions of years after core crossing.  

Given the distance the gas associated with the excess emission has traveled from its presumed former position coincident with the AGN (80 kpc), along with the bulk ICM velocities assumed necessary to disrupt a weak cool core (v$_{ICM}$ $>$ 1000 km s$^{-1}$), we can determine a maximum time since its displacement.  This value,  t$_{max}$ $\sim$ 8$\times$10$^7$ yr, is consistent with the value we calculate in $\S$ 6.2 for the synchrotron lifetime of the particles at the ends of the lobes of 0647+693.  This suggests that the WAT 0647+693 may have been ignited as the infalling cluster began to disrupt and displace the core.  As originally mentioned in \citet{Burns94} WATs may be triggered by mergers.  Of course, the merger ignition scenario may not hold true for all WATs, e.g. \citet{Sakelliou08}, but is likely a viable scenario in the case of A562.

\subsection{Radio Morphology and Galaxy Velocity Estimates}
As previously mentioned, a longstanding question remains over the extent to which WATs may be found in merger environments.  Estimates have been made to determine bulk flow velocities necessary to bend WATs into their observed shapes.  Using a heavy jet model and assuming \emph{in situ} particle reacceleration, \citet{Eilek84} and \citet{ODon93} determined that ICM velocities in excess of 1000 km s$^{-1}$ were required to shape the WATs.  As WAT host galaxies are generally the most massive members within their clusters which normally have low peculiar velocities, an ICM velocity greater than 200 km s$^{-1}$ could only be achieved during a merger.  The question was approached by \citet{Sakelliou96}, \citet{Hardcastle05}, and \citet{Jetha06} using a light jet model, excluding the assumption of \emph{in situ} particle reacceleration.  Their results suggest that only the most bent WATs require merger type ICM velocities to shape them, while those sources with unbent jets, but bent lobes, are more likely to be found in more relaxed systems with ICM velocities as low as $\sim$ 100 km s$^{-1}$.  Very little is understood about the internal density of the jets and lobes of WATs, resulting in such a broad range of possible ICM flow velocities.        

With A562 we have an opportunity to address the question from a slightly different angle.  Given that 0647+693 is clearly at the site of an ongoing merger and near the moment of core crossing, we can estimate an ICM flow velocity from previously published cluster merger simulations \citep{Roettiger99}.  From this, we may then work backwards to determine the WAT interior jet density that would result in the observed morphology.  Assuming the jets are bent simply due to the ram pressure of the ICM, we can apply Euler's equation:                            	 

\begin{equation}       
\frac{\rho_{r}v^{2}_{r}}{r_{c}}=\frac{\rho_{\rm ICM}v^{2}_{g}}{r_{r}},     
\end{equation}    

\noindent where $r_{r}$ is the radius of the jet, $\rho_{r}$ is its mass density, $v_{r}$ is the velocity of the plasma within the jet, $r_{c}$ is its radius of curvature, $v_{g}$ is the velocity of the host galaxy relative to the ICM, and $\rho_{\rm ICM}$ is the density of the ICM.  From the 8.4 GHz observation of the source (Fig. 10) we determine $r_{r}$=2 kpc and $r_{c}$=50 kpc.  By observing jet-sidedness within a sample of 30 WAT sources \citet{Jetha06} estimated a range of (0.3-0.7)c for jet speeds.  We adopt the average value of 0.5c.  From this \emph{Chandra} observation of A562, we have determined the ICM density within the central 50 kpc to be $\rho_{\rm ICM}$ = 4.5 $\times$ 10$^{-27}$ gm cm$^{-3}$.  Inserting these values and solving for $\rho_{r}$, we obtain an internal jet density of $\rho_{r}$ = 4.9 $\times 10^{-30}$ gm cm$^{-3}$ or $\rho_{r}$ $\sim$ 10$^{-3}\rho_{\rm ICM}$.  This falls within the range previously used by \citet{jetha06} and similar to values found by \citet{Smolcic07} with the z=0.22 source CWAT-01.            

A question of whether \emph{in situ} particle reacceleration occurs within the lobes of WATs presents an additional challenge in determining the speeds required to bend WAT sources.  If it is assumed that the radio luminosity of the source is supplied by a fraction of the kinetic energy of the particles flowing down the lobe, the bending of the lobes can be used to constrain the conditions surrounding the WAT/ICM interaction \citep{Eilek84}.  To determine if this is a valid assumption for 0647+693, we created a spectral index map of the radio emission between the frequencies of 1.4 and 8.4 GHz, shown in Fig. 11.  The spectral index, $\alpha$, of the northern hotspot is 0.7, where the radio flux $\rm S_\nu \propto \nu^{-\alpha}$.  Beyond the hotspot $\alpha$ remains relatively flat with increasing distance ($\alpha \sim$ 1.0), suggesting that reacceleration may be occurring to counteract spectral aging.    

If this interpretation is correct, and there is in fact \emph{in situ} particle reacceleration occurring within the lobes, then we may use the luminosity  condition \citep{Eilek84} combined with Euler's equation (Eq. 2) to calculate plasma velocities within the lobes:                        

\begin{equation} 
v_{r}=\frac{2 L_{\rm rad}}{\epsilon \pi \rho_{\rm ICM} v_g^2 r_c r_r},       
\end{equation}
\\

\noindent where $v_{r}$ is the plasma flow speed within the lobes, L$_{\rm rad}$ is the luminosity of the lobes, and $\epsilon$ is the radiative efficiency of the synchrotron plasma.  This efficiency term is poorly understood and a range of reasonable values (0.001-0.1) \citep{Birzan04} results in possible plasma flow velocities and densities spanning two orders of magnitude.  Such an uncertainty results in difficulties determining whether the bending of WAT lobes requires large merger induced velocities of the ICM (v$_{g}$ $>$ 1000 km s$^{-1}$) or more modest velocities caused by galactic orbital motion (v$_{g}$ $<$ 100 km s$^{-1}$).  We may estimate $\epsilon$ following the method of \citet{Birzan04}, where they examined 16 galaxy clusters which have clear decrements in the X-ray emission coincident with the lobes of cluster center radio sources.  It is assumed that this decrement in X-ray emission is a result of the relativistic plasma of the radio sources carving out cavities in the ICM.  The mechanical energy necessary to evacuate the cavities can be estimated from the product of the cavity volume (V) and overlying ICM pressure (P).  Assuming a standard time scale over which the cavity is created ($\sim 5 \times 10^7$ yr), a kinetic luminosity can then be estimated (L$_{\rm kin}$= PV/t$_{\rm rad}$).  The efficiency ($\epsilon$) is the ratio of total radio luminosity to kinetic luminosity.  As can be seen in Fig. 3, the northern radio lobe is clearly carving out a cavity within the intracluster gas.  The volume of the cavity, assuming a cylindrical lobe, is $\sim$ 8 $\times$ 10$^{69}$ cm$^3$, and the ICM pressure at this radius is 2.1 $\times$ 10$^{-11}$ dyn cm$^{-2}$.  This results in a total kinetic luminosity of 7 $\times$ 10$^{43}$ ergs s$^{-1}$.  We determine the total radio luminosity of the northern lobe to be 3.03 $\times$ 10$^{42}$ ergs s$^{-1}$, with a resulting radiative efficiency of $\sim$ 0.04.  This value is similar to that obtained for the WAT source 1159+583 in Abell 1446 \citep{Douglass08}.

The radius of the lobe at this distance from the cluster core is $\sim$ 26 kpc, with a radius of curvature of 25 kpc.  Assuming a bulk ICM flow of 1000 km s$^{-1}$, we determine the flow within the lobes is $v_{lobe} \sim$ 0.03c.  This is consistent with the values obtained in \citet{Smolcic07}. 

If, on the other hand, \emph{in-situ} particle reacceleration is not occurring within the lobes, we may estimate the flow velocity within the lobes by comparing the spectral age of the particles to the distance traveled since last acceleration (hotspots). To determine the spectral age of the plasma at various positions along the northern and southern lobes, fluxes were extracted from the 1.4 and 8.4 GHz images from which the average spectral indices were calculated.  Following the notation of \citet{Miley80}, we assume that the emission from the different regions comes from uniform prolate ellipsoids with a filling factor of unity and that there is equal energy in relativistic ions and electrons ($\kappa$ = 1). We also assume that the magnetic field is perpendicular to the line of sight and the spectral index is constant between the lower cutoff frequency of 10 MHz and the upper frequency cutoff of 100 GHz. Using these parameters for regions at the end of the lobes we determine the spectral ages to be t$_{spec}$ $\sim$ 3.0 $\times$ 10$^7$ yr, with a minimum energy magnetic field strength B$_{me}$ $\sim$ 4.6 $\mu$G for the northern lobe and t$_{spec}$ $\sim$ 2.9 $\times$ 10$^7$ yr, B$_{me}$ $\sim$ 5.0 $\mu$G for the southern lobe.  To travel the distance from the hotspot to the end of the southern lobe (D $\sim$ 170 kpc) within 30 Myr requires the the flow velocity within the lobe to be $v_{lobe} \sim$ 0.02c.  The agreement of this value with our luminosity condition estimate suggests that whether or not \emph{in situ} reacceleration is taking place, the particles are flowing down the lobes at 0.02c -0.03c.

Finally, to test our initial assumption that the ICM bulk flow velocity relative to the WAT host galaxy is on the order v$_g$ $\sim$ 1000 km s$^{-1}$, we use the buoyancy turn over method of \citet{Sakelliou96}.  As can be seen in the northern lobe of 0647+693 (Figs. 3, 10), there are three turnover points where the lobe changes its direction.  The first turnover, just past the hotspot, is presumably due to the bulk flow of the ICM bending the lobes.  The second turnover point, 60 kpc downstream,  is likely where the buoyancy of the lobe balances the external ram pressure of the ICM flow, beyond which the lobe bends radially outward.  The third turnover, 90 kpc further downstream is most likely due to merger induced eddies in the gas at larger radii \citep{Roettiger99}.  We can use the second turnover point, where buoyancy balances ram pressure, to determine the ICM flow velocity.  We use Eq. 11 from \citet{Sakelliou96}:

\begin{equation} 
v_{g}=\sqrt[]{1-\frac{\rho_{r}}{\rho_{ICM}} \frac{h k T}{\mu m_p} \frac{\nabla \rho}{\rho_{ICM}}}      
\end{equation}
\\

\noindent where $\rho_{r}$ and $\rho_{ICM}$ are the mass densities of the radio lobe plasma and the ICM, h is the width of the lobe at the turnover point, T is the local ICM temperature, $\mu$ is the mean molecular weight (0.6), and $\nabla \rho$ is the ICM density gradient at the turnover point ( 1.2 $\times$ 10$^{-26}$ g cm$^{-2}$).  We calculate a galaxy velocity relative to the ICM of 1190 km s$^{-1}$, consistent with the values expected within the cores of merging clusters \citep{Roettiger96}.

\section{Conclusions}
We have examined the galaxy cluster Abell 562, host to the prototypical merger-bent WAT source 0647+693.  \emph{Chandra} spectra of the source reveal a relatively constant temperature with radius, indicating the absence of an active cooling flow.  An elongation of the surface brightness distribution along the line that bisects the WAT is consistent with an ongoing merger occurring parallel to this line, resulting in the bent morphology of 0647+693.  A two-dimensional temperature map of the central 250 kpc reveals a region of increased temperature just east of the AGN, along the proposed merger axis.  The hotter region is likely the result of merger induced shock heating, possibly a reverse shock launched outward from the cluster center at the moment of core crossing. An image of the cluster, from which a smoothed $\beta$-model has been subtracted, reveals excess emission between the WAT lobes, which is believed to be gas displaced during the interaction. A two-dimensional abundance map shows a significant enhancement coincident with this excess emission.  Such a feature is possible evidence of a displaced weak cool core disrupted and heated by the merger.  The time since core displacement was determined from the location of the excess and the velocities presumed necessary to disrupt a weak cooling flow.  This value is of the order of the synchrotron lifetime determined for the plasma near the ends of the WAT lobes.  From this we may assume that it is possible that the WAT, 0647+693, was ignited by the merger itself.  

By combining observable properties of A562 and 0647+693 with common values for the conditions within merging clusters at the time of core crossing, we are able to constrain the internal density of the jets and plasma flow velocities within the lobes of the WAT.  We find the source is best represented by the model where $\rho_j$=0.001 $\rho_{ICM}$ with relatively slow (v$_r$ = 0.02c - 0.03c) plasma speeds in the lobes. Using the buoyancy turnover method of \citet{Sakelliou96} we confirm that ICM bulk flow velocities within the core of A562 are in excess of 1000 km s$^{-1}$.  

Support for this work was provided by the National Aeronautics and Space Administration, through Chandra Award Number GO6-7117X.  Basic research in radio astronomy at the Naval Research Laboratory is funded by 6.1 Base funding.  SWR was supported by the Chandra X-ray Center through NASA contract NAS8-03060, and the Smithsonian Institution. We thank John Houck and Josh Kempner for the S-Lang code and program to make the temperature map.

\bibliography{apjmnemonic,references}
\bibliographystyle{apj}

\clearpage

    \begin{figure}
\epsscale{1.1}
      \plotone{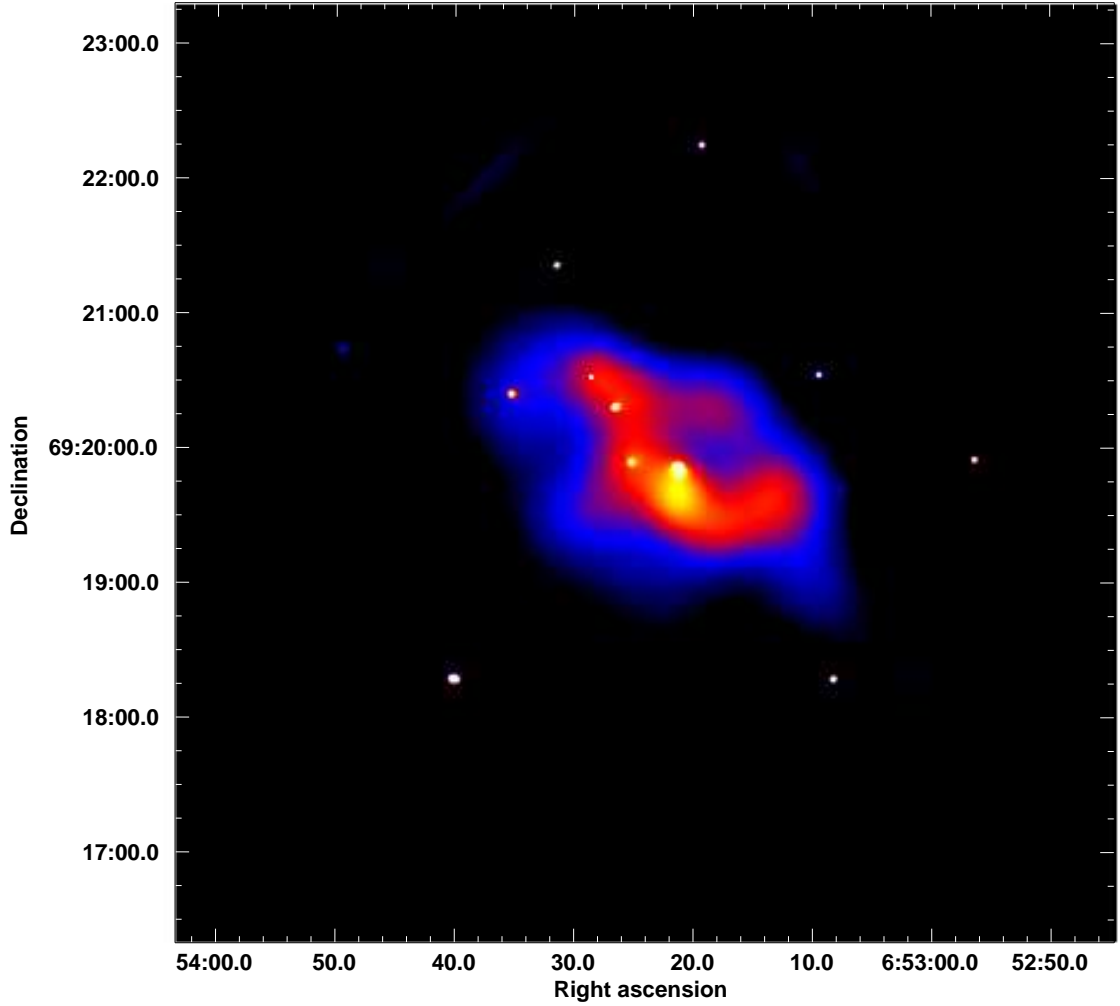}
      \caption{\label{fig:sed} An adaptively smoothed Chandra ACIS-S3 image of the 600 $\times$ 600 kpc$^2$ region of the galaxy cluster Abell 562.  The image has been corrected for background and exposure.  The cluster appears to be the site of a merger occurring parallel to the northeast-southwest elongation of the emission.}

 \end{figure}

    \begin{figure}
	\epsscale{0.85}
      \plotone{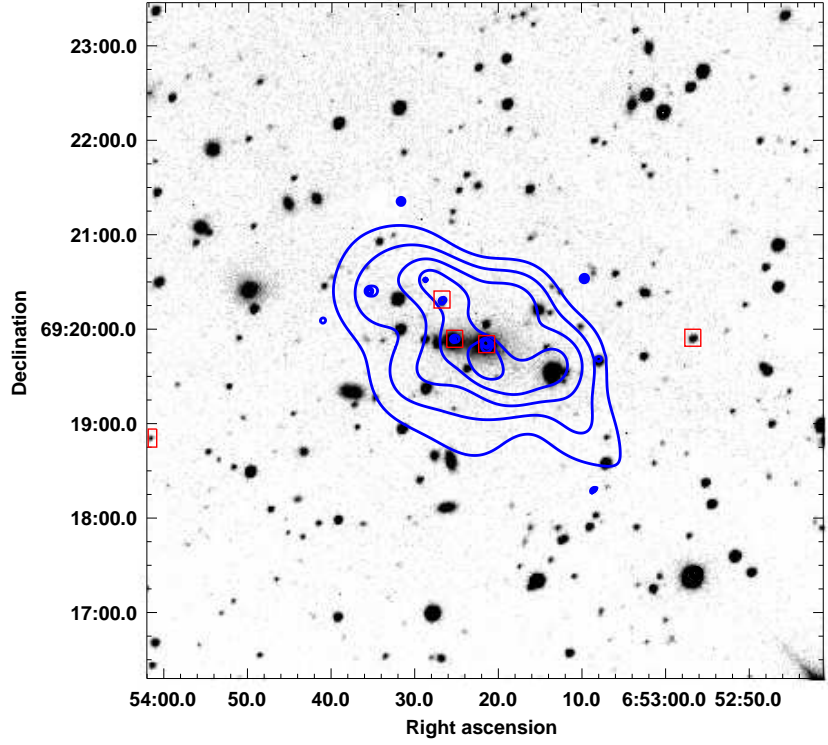}
      \caption{\label{fig:sed} Perkins 72" R-band image of Abell 562 with overlays of X-ray contours of adaptively smoothed emission obtained with Chandra (Figure 1).  The contours are logarithmically spaced between 0.68 cts arcsec$^{-2}$ and 2.8 cts arcsec$^{-2}$. Red squares indicate the point sources identified using the wavdetect algorithm that are coincident with optical sources in the USNO A2.0 catalog.}

    \end{figure}

    \begin{figure}
\epsscale{0.9}
      \plotone{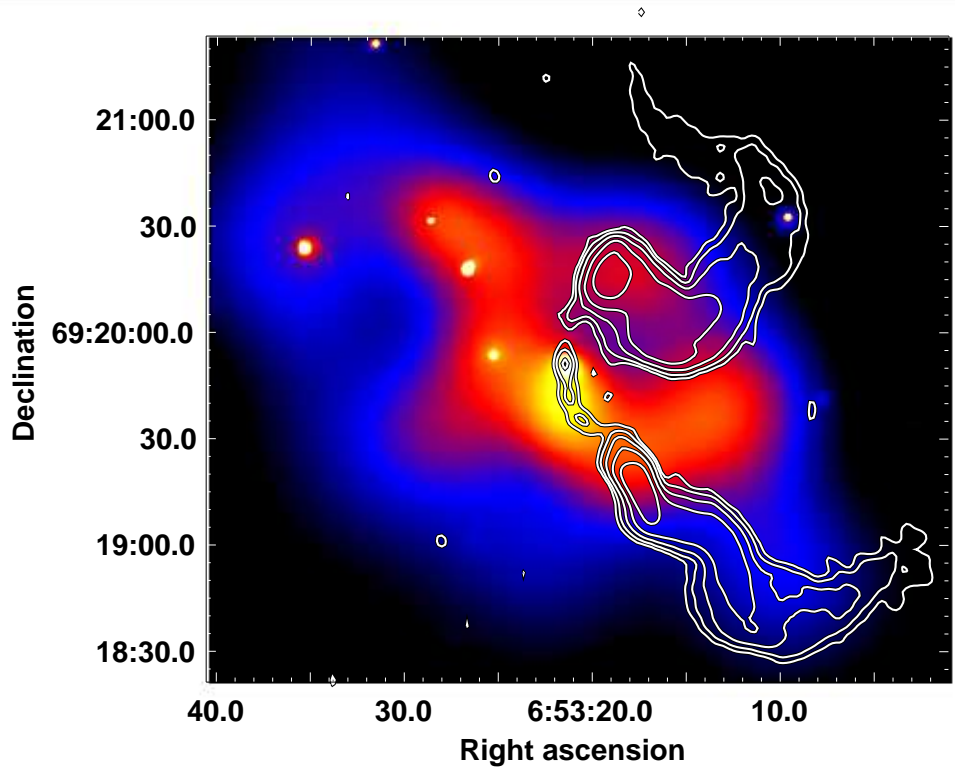}
      \caption{\label{fig:sed}Adaptively smoothed \emph{Chandra} image of A562.  Contours of the 1.4 GHz VLA map of 0647+693 are overlaid in white.  The cluster emission is elongated parallel to the line bisecting the WAT.  Additionally, the northern lobe is coincident with a decrement in emission likely due to the radio source having carved out a cavity within the cluster gas.}

 \end{figure}

    \begin{figure}
      \plotone{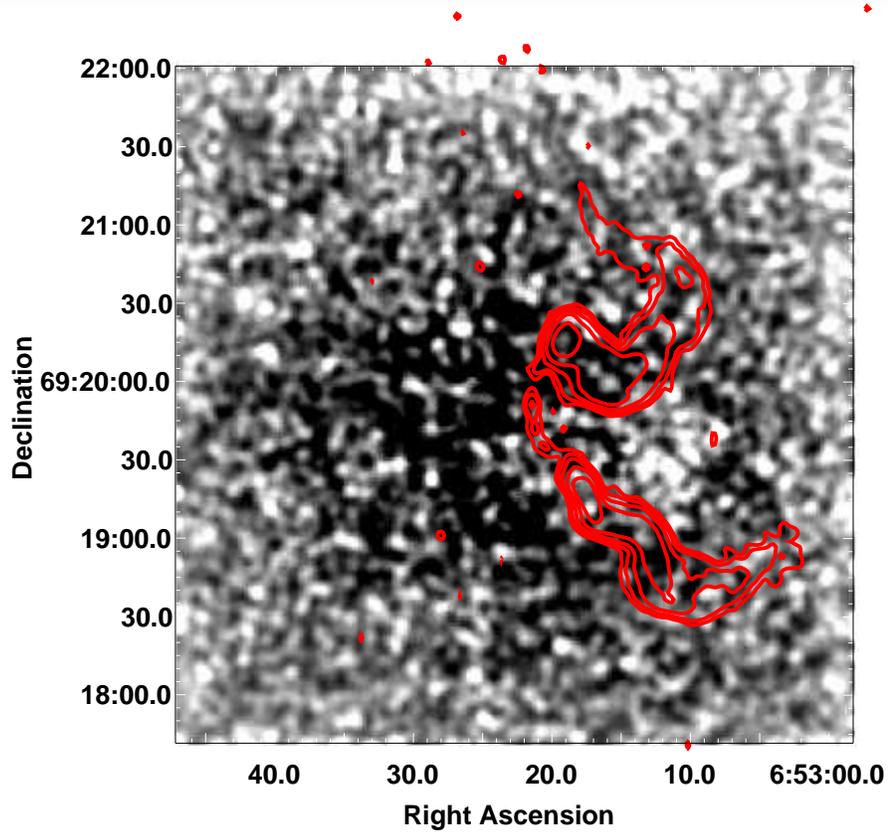}
      \caption{\label{fig:sed} An image of excess X-ray emission created by subtracting a two-dimensional $\beta$-model from a 4$\arcsec$ Gaussian smoothed image of the cluster.   Radio contours are the same as in Fig. 3. An excess of emission can be seen between the radio lobes.  This is likely displaced gas by ram pressure from an infalling sub cluster.}

    \end{figure}

   \begin{figure}
      \plotone{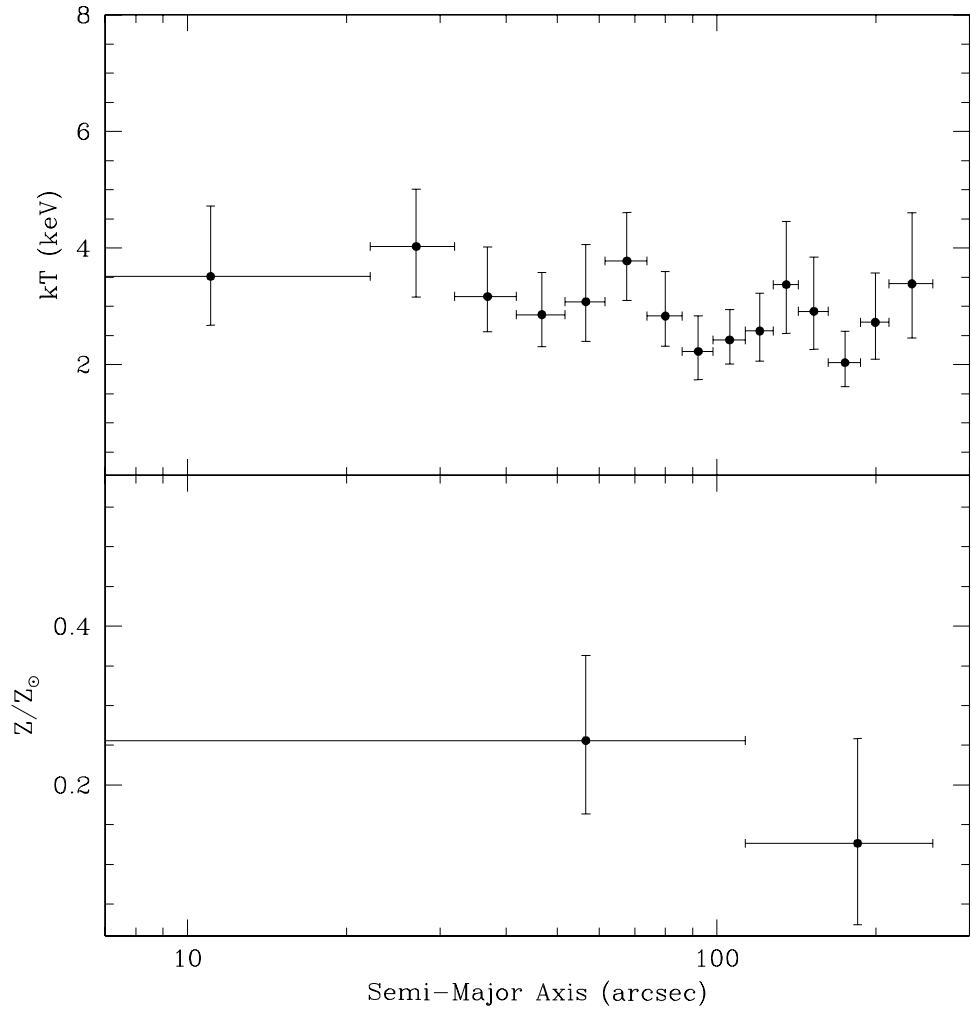}
      \caption{\label{fig:sed} Profiles of temperature and chemical abundance.  The temperature profile is essentially flat out to a radius of 200$\arcsec$ (400 kpc).  The abundance values hint at a slight decrease with increasing radius, although this decrease is not statistically significant.}
    \end{figure}

    \begin{figure}
	\epsscale{1.0}
      \plottwo{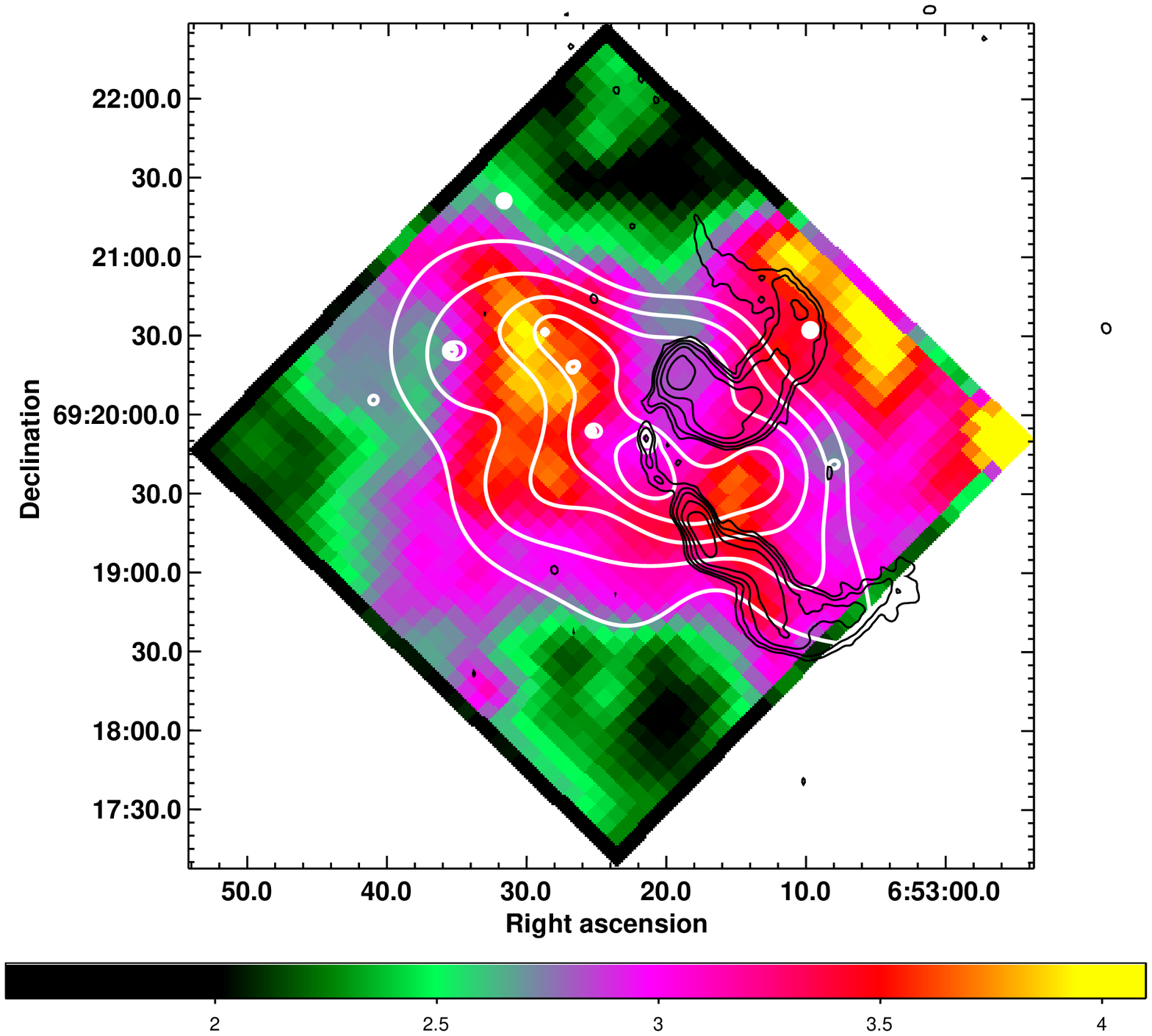}{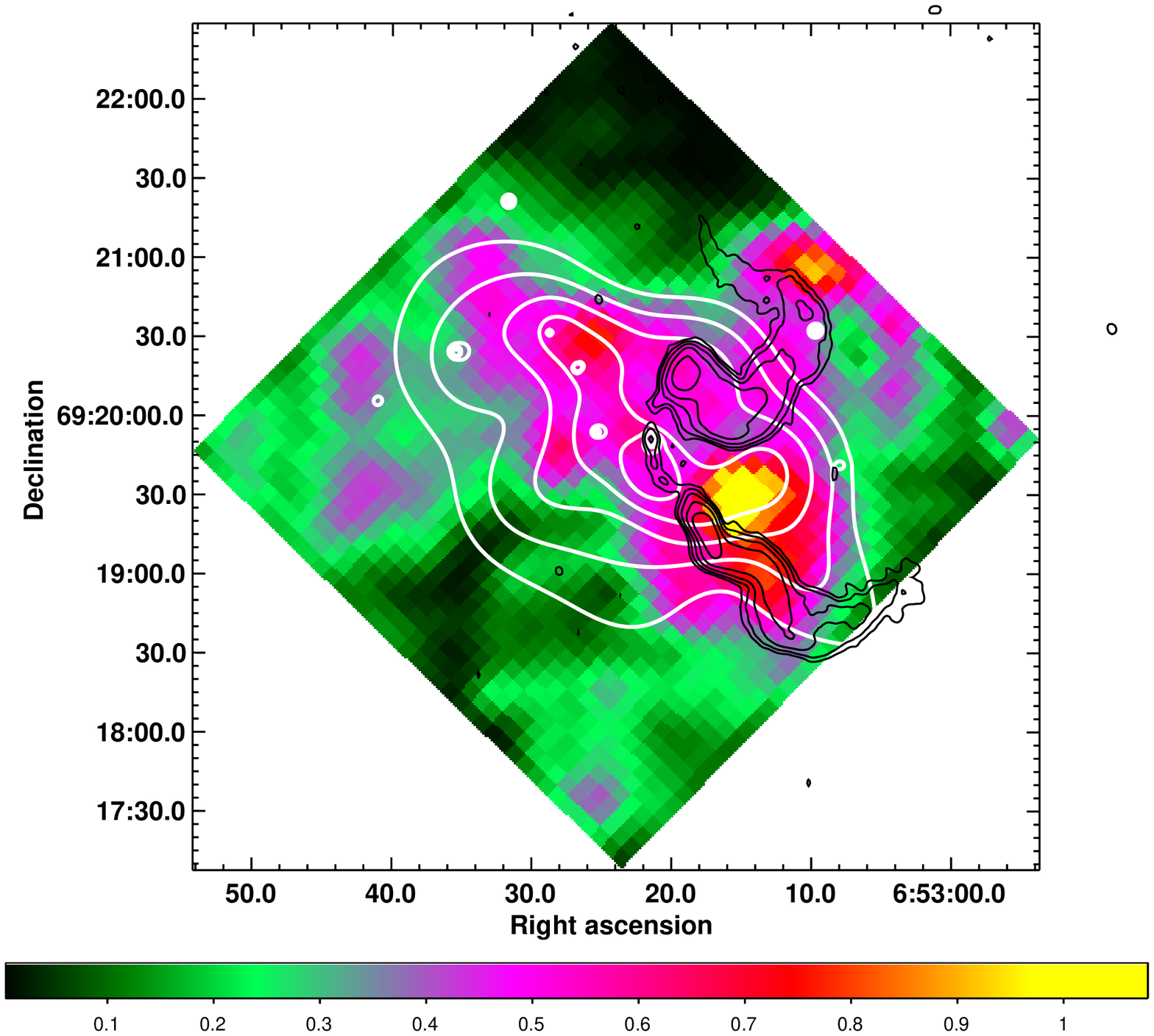}
      \caption{\label{fig:sed} Temperature (left) and abundance (right) map of the cluster center with contours of the smoothed X-ray (white) and radio (black) superposed.  In the temperature map, a region of significantly hotter emission can be seen directly to the east of the central AGN.  This is most likely gas heated by a reverse shock induced by the merger.  In the abundance map a region of higher metallicity is seen between the lobes.  It is coincident with the X-ray excess seen in Fig. 4, and it may indicate the presence of a disrupted cool core. }

    \end{figure}

    \begin{figure}
	\epsscale{0.85}
      \plotone{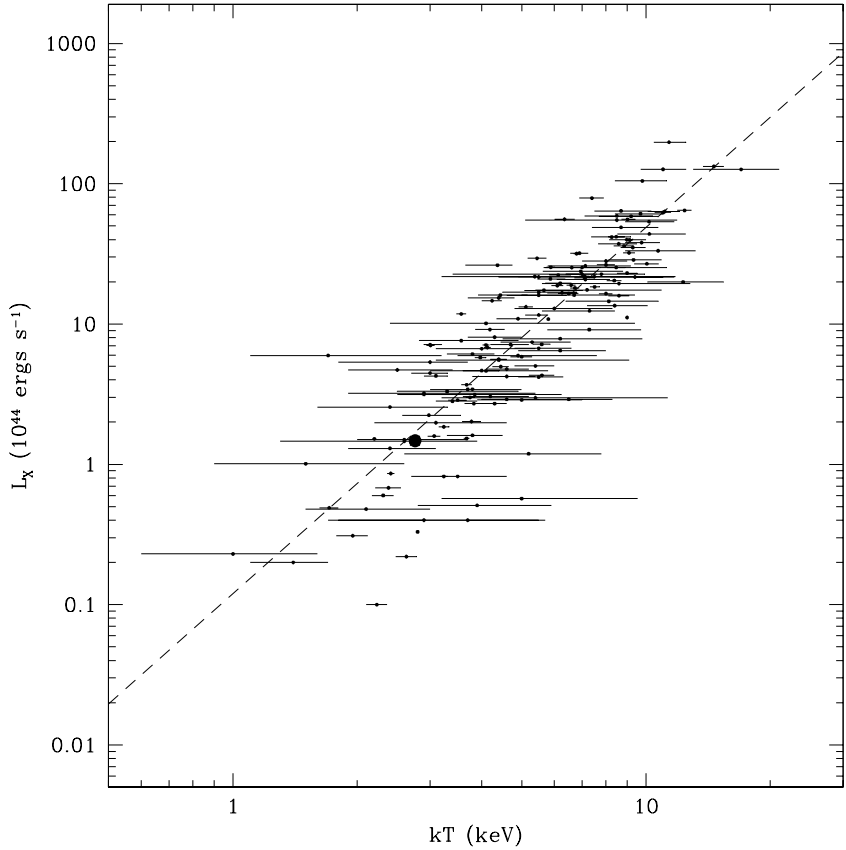}
      \caption{\label{fig:sed} $L_X-T$ relation plot from \citet{Wu99}.  Abell 562 is denoted by the large solid point.  The relation appears to fall along the trend. }

    \end{figure}

   \begin{figure}
      \plotone{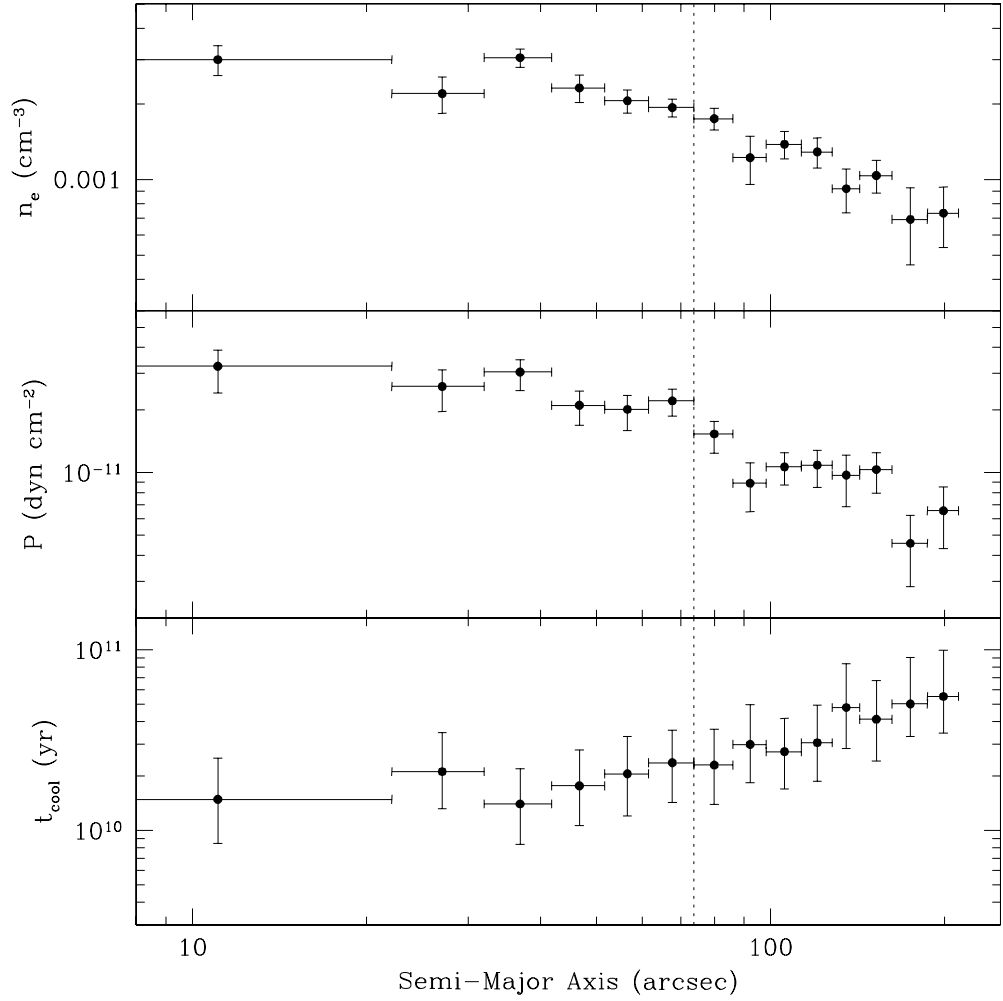}
      \caption{\label{fig:sed} Profiles of electron number density, pressure, and cooling time. The dotted line indicates the approximate radius of the pressure jump.}
    \end{figure}

   \begin{figure}
   \epsscale{0.75}
      \plotone{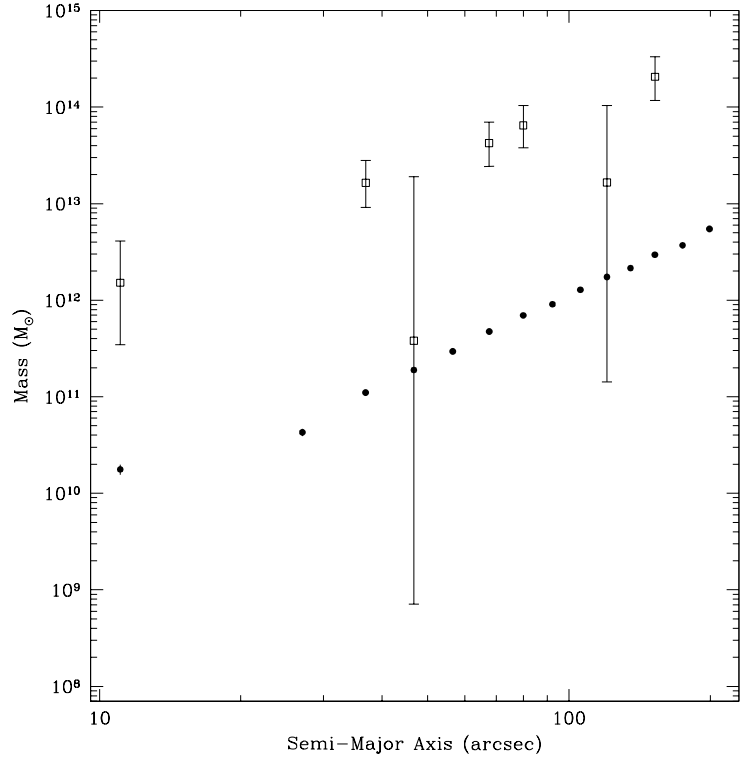}
      \caption{\label{fig:sed} Profiles of gas (circles) and gravitational (squares) mass. As the radius increases, the gas mass fraction is significantly lower than that in more relaxed clusters suggesting a failure of the assumption of hydrostatic equilibrium within the mass calculation.}
    \end{figure}

    \begin{figure}
	\epsscale{0.85}
      \plotone{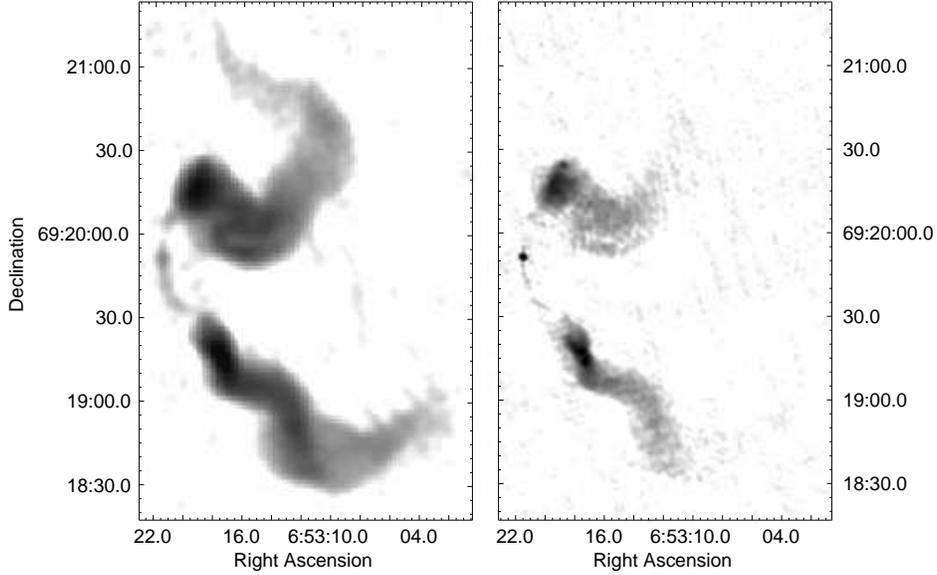}
      \caption{\label{fig:sed} VLA 1.4 GHz (left) and 8.4 GHz (right) maps of 0647+693 observed in the A and A+B+D configurations respectively.  Images show identical fields of view.}

    \end{figure}

    \begin{figure}
	\epsscale{0.85}
      \plotone{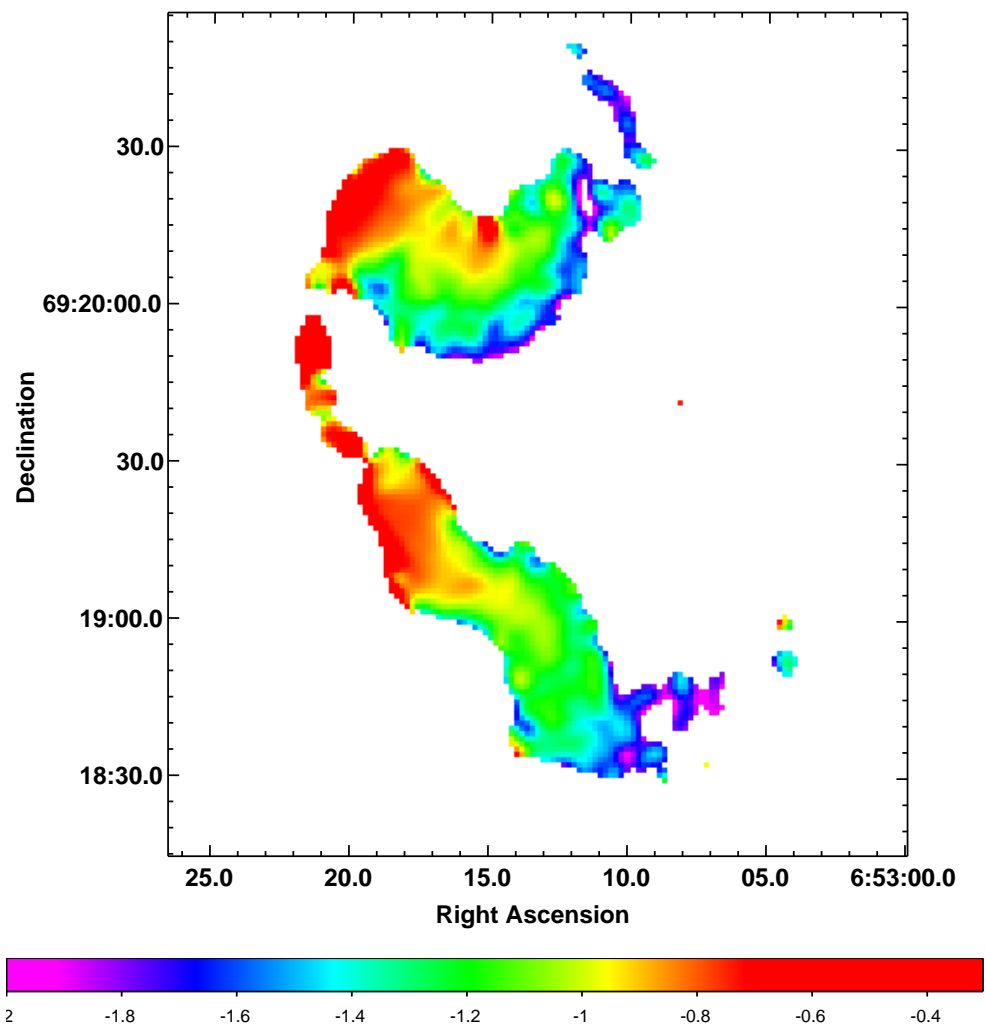}
      \caption{\label{fig:sed} Spectral index map of 0647+693 obtained from 1.4 GHz and 8.4 GHz VLA data.}

    \end{figure}

\begin{deluxetable}{lccccc}
\tablecaption{Radio Observations of 0647+693}
\tablewidth{0pt}
\tablehead{
\colhead{Date} & \colhead{VLA Configuration}   & \colhead{Frequency}   &
\colhead{Bandwidth} &
\colhead{Duration} &
\colhead{Obs.\ Code}\\
\colhead{} & \colhead{} & \colhead{(MHz)} & \colhead{(MHz)} & \colhead{(hours)} & \colhead{}
}
\startdata
1998 Jul 27 & A & 1465.0/1435.0 & 25/25 & 1.2 & AL445\\
2001 Jan 15 & A & 8447.6/8497.6 & 25/25 & 1.9 & AH717\\
2001 May 19 & B & 8435.1/8485.1 & 50/50 & 1.3 & AH717\\
2001 Dec 18 & D & 8435.1/8485.1 & 50/50 & 0.2 & AH717\\
2001 Dec 20 & D & 8435.1/8485.1 & 50/50 & 0.2 & AH717\\
\enddata
\label{tbl:radio}
\end{deluxetable}

\end{document}